\documentclass{IOS-Book-Article}

\usepackage{graphicx}
\usepackage{mathptmx}
\usepackage{soul}\setuldepth{article}
\usepackage{enumitem}
\usepackage{tabularx}
\usepackage{subcaption}
\usepackage{comment}
\def\hb{\hbox to 11.5 cm{}}

\usepackage[table]{xcolor}
\usepackage{soul}
\definecolor{tan}{rgb}{0.82, 0.71, 0.55}
\definecolor{silver}{rgb}{0.75, 0.75, 0.75}

\definecolor{keylime}{rgb}{0.70, 1.00, 0.40}
\definecolor{electricblue}{rgb}{0.33, 0.71, 0.98}
\definecolor{lightpurple}{rgb}{0.75, 0.58, 0.89}
\definecolor{seafoam}{rgb}{0.47, 0.87, 0.78}
\definecolor{brightmint}{rgb}{0.53, 0.98, 0.65}
\definecolor{lightpurple}{rgb}{0.75, 0.58, 0.89}
\definecolor{lavender}{rgb}{0.90, 0.90, 0.98}
\definecolor{softmint}{rgb}{0.78, 0.98, 0.69}
\definecolor{lightpeach}{rgb}{0.99, 0.90, 0.82}
\definecolor{pastellavender}{rgb}{0.90, 0.87, 0.98}
\definecolor{skyblue}{rgb}{0.69, 0.87, 0.93}
\definecolor{palebutter}{rgb}{0.99, 0.98, 0.76}
\definecolor{warmpink}{rgb}{0.98, 0.82, 0.86}
\definecolor{powderblue}{rgb}{0.82, 0.90, 0.96}
\definecolor{pistachio}{rgb}{0.88, 0.96, 0.78}
\definecolor{maize}{rgb}{0.98, 0.93, 0.36} 
\definecolor{salmon}{rgb}{0.98, 0.60, 0.60} 
\definecolor{lightbrown}{rgb}{0.91, 0.82, 0.72}

\definecolor{coral}{rgb}{1.00, 0.50, 0.31} \definecolor{hotpink}{rgb}{1.00, 0.41, 0.71}  
\begin{document}

\pagestyle{headings}
\def\thepage{}
\begin{frontmatter}              

\title{Uncovering AI Governance Themes in EU Policies using BERTopic and Thematic Analysis }

\markboth{}{\hb}

\author[A]{\fnms{Delaram} \snm{Golpayegani}\orcid{0000-0002-1208-186X}%
\thanks{Corresponding Author:Delaram Golpayegani, golpayes@tcd.ie.}},
\author[A, B]{\fnms{Marta} \snm{Lasek-Markey}\orcid{0000-0002-0183-3982}},
\author[C]{\fnms{Arjumand} \snm{Younus}\orcid{0000-0001-7748-2050
}},
\author[A, C]{\fnms{Aphra} \snm{Kerr}\orcid{0000-0001-5445-7805}}, and
\author[A, B]{\fnms{Dave} \snm{Lewis}\orcid{0000-0002-3503-4644}},

\runningauthor{D. Golpayegani et al.}
\address[A]{ADAPT Centre}
\address[B]{Trinity College Dublin, Dublin, Ireland}
\address[C]{University College Dublin, Dublin, Ireland}

\begin{abstract}
The upsurge of policies and guidelines that aim to ensure Artificial Intelligence (AI) systems are safe and trustworthy has led to a fragmented landscape of AI governance. The European Union (EU) is a key actor in the development of such policies and guidelines. Its High-Level Expert Group (HLEG) issued an influential set of guidelines for trustworthy AI, followed in 2024 by the adoption of the EU AI Act. While the EU policies and guidelines are expected to be aligned, they may differ in their scope, areas of emphasis, degrees of normativity, and priorities in relation to AI.
To gain a broad understanding of AI governance from the EU perspective, we leverage qualitative thematic analysis approaches to uncover prevalent themes in key EU documents, including the AI Act and the HLEG Ethics Guidelines. We further employ quantitative topic modelling approaches, specifically through the use of the BERTopic model, to enhance the results and increase the document sample to include EU AI policy documents published post-2018. We present a novel perspective on EU policies, tracking the evolution of its approach to addressing AI governance. 
\end{abstract}

\end{frontmatter}
\markboth{September 2025\hb}{September 2025\hb}

\section{Introduction}

The rapid advancement of Artificial Intelligence (AI) technologies, coupled with an increasing number of AI-related incidents, has positioned AI as a key topic in global policymaking discourse. The European Union (EU) has become a leader in this landscape, particularly after the enactment of its AI regulation, the EU AI Act~\cite{eu-ai-act}
, in 2024. However, multiple documents were issued both before and after the adoption of the AI Act that have shaped the discourse in the EU. These range from the EU-High Level Expert Group (HLEG)'s ``Ethics Guidelines for Trustworthy AI'' (2019)~\cite{eu-hleg-trustworthyai}, to the guidelines issued by the European Commission in 2025 on prohibited AI practices (C(2025) 5052 final)~\cite{eu_prohibited}. While EU AI policies are expected to be aligned in pursuit of a common vision of trustworthy AI, they adopt distinct approaches towards AI governance, which is manifested in their focus area (e.g. AI innovation, risk management, sector-specific requirements), their degree of normativity (binding, non-binding), enforcement mechanisms, and technical and organisational measures they prescribe. There is also a temporal dimension to consider, since these policies were developed at different times and therefore  may not address recent advances in AI capabilities. 

 

In this paper, we aim to uncover AI governance concepts within the EU AI policy landscape through identification of recurring topics and themes, both shared and divergent. The research question this paper explores is: \textit{What are the prevalent themes in the EU AI policies and guidelines post-2018 concerning AI governance?}

One way of addressing the research question is through the use of \textit{qualitative thematic analysis} approaches~\cite{Clarke2014thematic} from social sciences. Since thematic analysis is conducted manually by domain experts, it can yield nuanced and high-quality insights, however it is inherently subjective, time-intensive, and difficult to scale. To mitigate these limitations, we adopt \textit{unsupervised topic modelling} techniques from natural language processing (NLP), which enable identification of frequent topics in large text corpora, in combination with thematic analysis to interpret, validate, and provide context to the discovered topics.

In the remainder of this paper, we first review the body of work on the analysis of AI policies in \autoref{sec:<related_work>}. Then, in \autoref{sec:<methodology>} we discuss our methodology that combines topic modelling and thematic analysis approaches.  \autoref{sec:<topic_modelling>} provides the identified AI topics and themes.  
Finally, \autoref{sec:<conclusion>} concludes the paper by discussing our plan for future work. 

\section{Related Work} \label{sec:<related_work>}

There is a body of work that analyses the content of AI policies and guidelines to determine the degree of consensus (or divergence) as well as the gaps in AI governance and ethical AI principles within the growing body of AI policies. These studies apply qualitative thematic analysis, quantitative content analysis (often using topic modelling algorithms such as LDA and BERT), or a combination of both. In the following, we review some of the related work on AI policies that utilised these approaches.


\subsection{Automated Analysis of AI Policies via Topic Modelling} 
A pioneering work \cite{theodoros2020} within the realm of NLP-based AI policies' analysis uses topical distributions from LDA to perform clustering over 12 national AI strategies, and use the cluster information to determine the strategic priorities of various governments in the AI race. It is important to note that this analysis was performed over some of the earliest available documents covering strategic directions laid by national governments with respect to AI; and hence there is also more coverage on leadership in AI deployment for innovation.
Another work built on LDA-based analysis and visualisations mines China's data governance policies to discover evolution paths within these \cite{yang2022lda}. One of the uncovered themes is AI-based decision-making for data governance. Similarly, Wang et al. \cite{wang2025artificial} pursue via structured topic models a comparative analysis of AI policies from China, EU and US uncovering significant differences in priority areas within the AI policymaking landscape.

\subsection{Combination of Automated and Manual Analysis of AI policies}
Roche et al.~\cite{roche2023ethics} used a combination of automated data analytics tools and manual content analysis on a corpus of AI policies and governance frameworks to investigate the inclusivity of ethical AI approaches. Specifically, inclusivity here focuses on the contribution of voices from the Global South within the AI ethics discourse. Schiff et al. ~\cite{schiff2021aiethics} analysed the degree of global consensus on ethical AI principles by manual coding of AI documents published by public sector, private organisations, and NGOs, followed by a quantitative analysis of the coding. 
Works by Saheb and Saheb ~\cite{saheb2023topical,saheb2024mapping} perform a mixed method analysis of AI ethics guidelines issued by national governments to identify the most pressing concerns with respect to the ethical aspects of AI. Specifically, their methodology involves the application of topic modelling\footnote{They use the LDA variant embedded in the tool WordStat.} followed by manual extraction of themes from within the documents. A very recent work by Kajava et al. \cite{Kajava2025postgpt} attempts to track the discourse surrounding increased risks in AI landscape following the launch of ChatGPT. Their method comprises the use of guided BERTopic in conjunction with qualitative content analysis; and this was done to ensure scalability and trust in the analysis of the discourse. Similarly, Suter et al. \cite{suter2025politicians} track AI discussions in various parliamentary debates through a combination of BERTopic and manual classification into interpretable themes by two independent coders. They specifically analyse debates from US Congress, EU Parliament, Parliament of Singapore, and Swiss Federal Assembly and this helps them uncover distinct national priorities in a post-LLM world. Essentially, the findings of both works \cite{Kajava2025postgpt,suter2025politicians} reveal a greater focus on increasing literacy and ethical awareness within the adoption of AI policies in a world that has been shaken by software such as ChatGPT.

A key difference between our analysis and the existing ones lies in the way we track the progress made by a single entity. i.e., the European Union, which has emerged as a leader in AI governance and regulatory landscape. On a superficial level, this may seem similar to the analysis by Kajava et al. \cite{Kajava2025postgpt}. Still, their primary focus is on tracking changes in risk discourse within EU policy circles following the launch of ChatGPT. Conversely, ours seeks to understand the discourse as a whole, tracking overall continuities and shifts over time. This is especially important as implementation of the AI Act is dependent on the development of guidelines, codes of conduct, codes of practice and technical standards. The Act is also subject to changes via implementing and delegated acts as well as revisions arising from the EU Omnibus Programme that aims to minimise needless duplication of compliance and oversight effort. This presents a complex and changing body of texts that must be monitored and analysed to understand changes to the expectations and effect of the original Act. Moreover, our analysis is complemented by a comparison conducted by experts from law, computer science, and social science, resulting in refined themes that effectively capture the complexities and nuances of EU policymaking efforts regarding AI governance. 






\section{Methodology} \label{sec:<methodology>}

\begin{figure} [!]
    \centering
    \fbox{\includegraphics[width=0.75\linewidth]{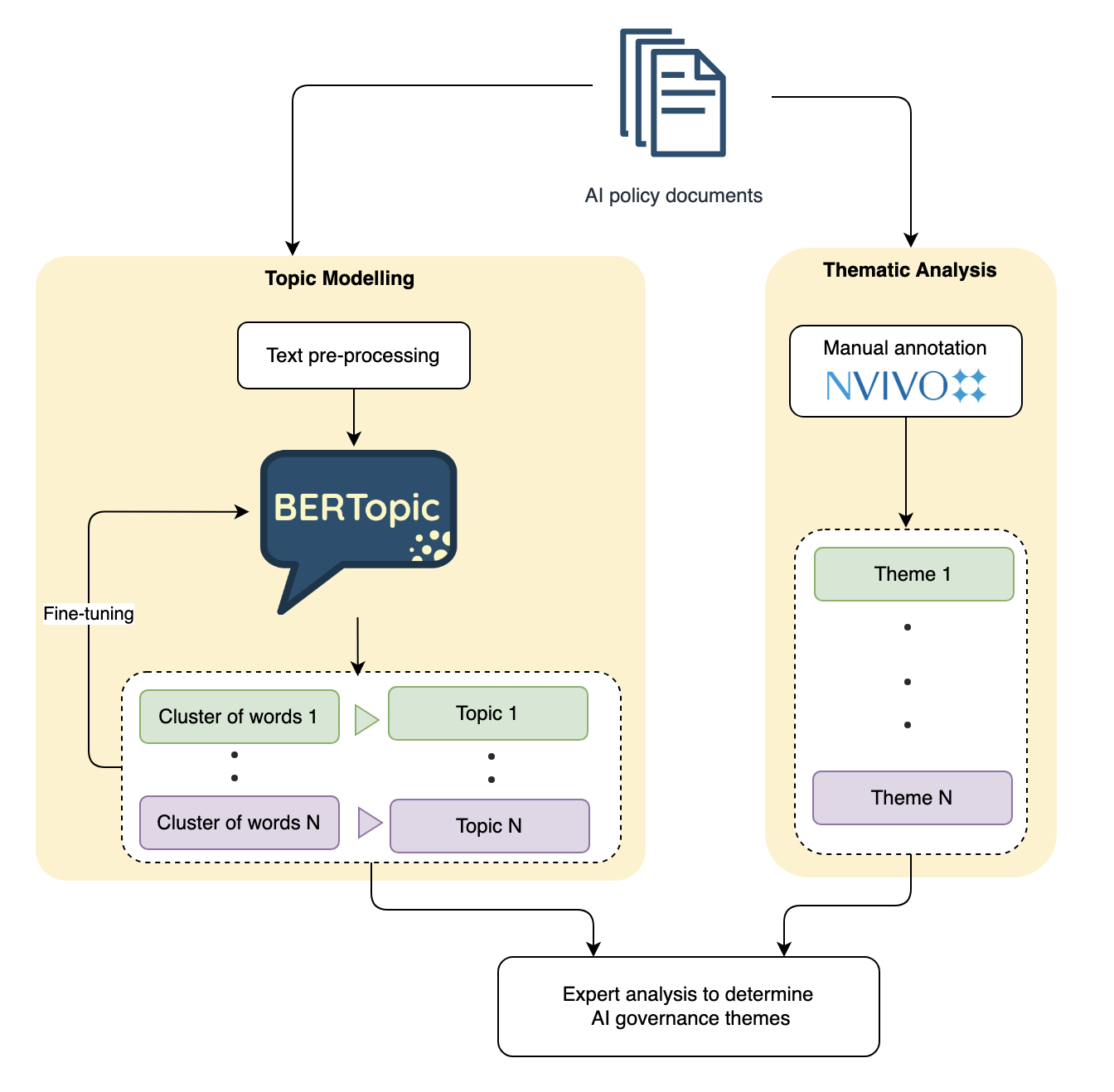}}
    \caption{The overall methodology}
    \label{fig:<methodology>}
\end{figure}

 As shown in \autoref{fig:<methodology>}, our methodology comprises four key steps: (1) corpus creation of EU AI policy documents, (2) manual thematic analysis, (3) unsupervised topic modelling and (4) interdisciplinary expert analysis of topics to identify themes and derive additional insights. For the purpose of this paper, we consider a \textit{topic} as a concept identified automatically by a topic modeller from text (algorithm-driven), and a \textit{theme} as a broad interpretable semantic concept derived through expert interpretation,  either from the topics generated by a topic modeller or directly from the text itself (human-driven). 

\subsection{Corpus Creation} 
Our study focuses on EU AI policy documents issued between January 2018 and August 2025, covering the period from early trustworthy AI guidelines through the recent regulation setting efforts. Our corpus includes legislation, official guidelines, and codes of practice from relevant EU institutions. Given that English is one of the official EU languages, all of the documents were published in English, which we used for our analysis. The list of documents that constitute our corpus is provided in  \autoref{tab:<corpus>}. 

\begin{table}[!t]
    \centering
    \caption{EU AI policy documents included in the corpus}
    \begin{tabularx}{\textwidth}{|p{0.4cm}|X|p{2cm}|p{2cm}|p{1cm}|}
\hline
    \textbf{ID} &
    \textbf{Document}  & \textbf{Issuer} &  \textbf{Type} & \textbf{Year} \\

\hline

 01 & Policy and investment recommendations for trustworthy AI \cite{eu_hleg-policy} & EU HLEG & Recommendation & 2019 \\
 
 \hline
    
   02 & Ethics guidelines for trustworthy AI \cite{eu-hleg-trustworthyai} & EU HLEG  & Guideline & 2019 \\

\hline
  03 &  Assessment List for Trustworthy AI (ALTAI) \cite{eu-hleg-altai} & EU HLEG & Guideline & 2020 \\
\hline   

  04 &   Sectoral considerations on policy and investment recommendations for trustworthy AI \cite{eu-hleg-sectoral} &  EU HLEG & Recommendation & 2020 \\

 \hline 

    05 &   EU AI Act \cite{eu-ai-act} & EU Council and Parliament & Legislation & 2024 \\
\hline 
    06 &  Commission Guidelines on prohibited AI practices established by the AI Act \cite{eu_prohibited} &  European Commission & Guideline & 2025 \\
\hline      
    07 &  Commission Guidelines on the scope of the obligations for general-purpose AI models established by the AI Act \cite{eu_genai_models}  &   European Commission  & Guideline &  2025 \\
\hline
    08 & EU Code of Practice for General-Purpose AI Models \cite{eu_gpai_code}, including \textit{transparency}, \textit{copyright}, and \textit{safety and security} chapters & European Commission & Code of practice & 2025 \\
\hline        
    \end{tabularx}
    
    \label{tab:<corpus>}
\end{table}

\subsection{Thematic Analysis}
Thematic analysis is a popular method of qualitative data analysis that systematically organises datasets, and helps to identify patterns of meaning commonly referred to as themes~\cite{squires2023thematic}. First described in the 1970s by Houlton – albeit the term itself had been in use even earlier – it became more prominent in the late 1990s with researchers like Boyatzis and Hayes~\cite{Clarke2014thematic}. In recent times, thematic analysis has been understood as an umbrella term for different approaches. While popular in qualitative interview data analysis, computer-assisted expert thematic analysis of legal texts, such as legislation or policy documents, appears to be less commonly employed.

We performed qualitative analysis of a subset of the corpus, including the EU AI Act and the HLEG ``Ethics Guidelines for Trustworthy AI'' with the use of NVivo software. As explained by Dhakal, NVivo is a CAQDAS programme that assists, rather than replaces a human researcher~\cite{dhakal2022nvivo}. NVivo analysis is, thus, expert-led and in this project it was performed by a legal researcher. The documents were imported into NVivo and thematic coding was performed by selecting a section of text and then tagging it with a node corresponding to a theme. Rather than being preassigned ahead of time, the nodes were created \textit{in situ}, as was deemed necessary by the human researcher. As a result, the EU AI Act analysis resulted in 59 themes, while the HLEG Guidelines resulted in 52 themes, with multiple themes overlapping across both documents. 

As mentioned above, in this project thematic analysis was conducted on a subset of documents in parallel to topic modelling. The outputs of two methods were then compared, with the results of the manual thematic analysis serving to validate the topic modelling approach.  

\subsection{Topic Modelling}

 Topic modelling refers to identification of prevalent topics in corpora of text using unsupervised machine learning techniques and is primarily used for uncovering topics within large sets of documents. However, this work only deals with limited number of documents that should be processed separately. Therefore, to enable applying a topic modelling algorithm on a single document, we treated each document as a set of smaller documents. For this, using the \texttt{PyMuPDF} library in Python, we extracted the text from the PDF file and then split the text into smaller parts, in this case sentences, and 
 stored them in a CSV file.





 There exist a variety of unsupervised topic models (for evolution of such models see~\cite{churchill2022evolution}). Among the existing topic models, in consultation with NLP experts, we initially considered Latent Dirichlet Allocation (LDA)~\cite{blei2003lda} and BerTopic~\cite{grootendorst2022bertopic}. We applied both topic modelling algorithms to a sample of policy documents (the EU AI Act, HLEG trustworthy AI guidelines, and UNESCO's guideline on the ethics of AI~\cite{unesco2022ethics}\footnote{ 
 Since the UNESCO recommendation is not an EU-specific policy, it is out of the scope of the work presented in this paper. However, we included it in this step as a comparative reference to ensure the generalisability of our approach across different policy contexts.}) and compared their outputs against the results of manual thematic analysis. Our findings indicate that BERTopic outperforms LDA in the context of AI policy texts, offering higher accuracy and better interpretability. This is aligned with the findings of Kajava et al. ~\cite{Kajava2025postgpt}, which compared the interpretability of these two topic modellers for EU-related documents and news articles on AI.

 Before application of \texttt{BERTopic}, we employed common NLP approaches to apply minimal preprocessing using the \texttt{nltk} library for: 
(1) lowercasing , (2) removing  non-alphabetic tokens (e.g. punctuation marks),
(3) removal of tokens that consist solely of numbers, and (4) exclusion of tokens shorter than two characters. We did \emph{not} remove stop words or perform lemmatisation prior to embedding, given that BERTopic uses transformer-based embeddings, which require the text to be persevered for better accuracy. We remove the stop words, after generating embeddings, using the \texttt{CountVectorizer} when initialising BERTopic.
 We fine-tuned \texttt{BERTopic} model in a iterative manner to ensure that the model identifies at least 3 topic categories for each document. The snippets used for preprocessing and topic modelling are available on GitHub under the MIT licence at \url{https://github.com/DelaramGlp/forsee_topicmodelling}.   


\subsection{Expert Analysis}

The expert analysis was conducted in two phases, each serving a different purpose:
\begin{enumerate}
    \item \textbf{Validation of topic model}: This phase aimed to identify the degree of topic modelling accuracy by comparing its outputs with the results of manual thematic analysis. An analysis of the HLEG trustworthy AI guidelines  and the EU AI Act as representative case studies was conducted by the authors, who are experts in law, computer science and social sciences. The results serve as a basis for applying automated analysis to the remaining documents in the corpus. 
    
    \item \textbf{Topic interpretation and theme discovery}: This phase involved qualitative refinement of the BERTopic results. The first and second authors of this paper, both experts in EU AI policies, independently analysed the generated topics and assigned themes to each topic category. The proposed themes were then discussed in a joint session and consolidated into their final form. 
    
    \end{enumerate}




\section{AI Governance Themes Identified in BERTopic Results} \label{sec:<topic_modelling>}

\subsection{From Topics to Themes}
Applying the BERTopic model to each document produced 227 topic clusters in total (see the breakdown of the topic clusters in \autoref{tab:<bert_topics_themes>}). The output of the topic modelling for each document is visualised using both bar charts and word cloud digrams, and the corresponding results are stored in CSV format. 

\begin{figure}[!h]
    \centering \fbox{\includegraphics[width=\linewidth]{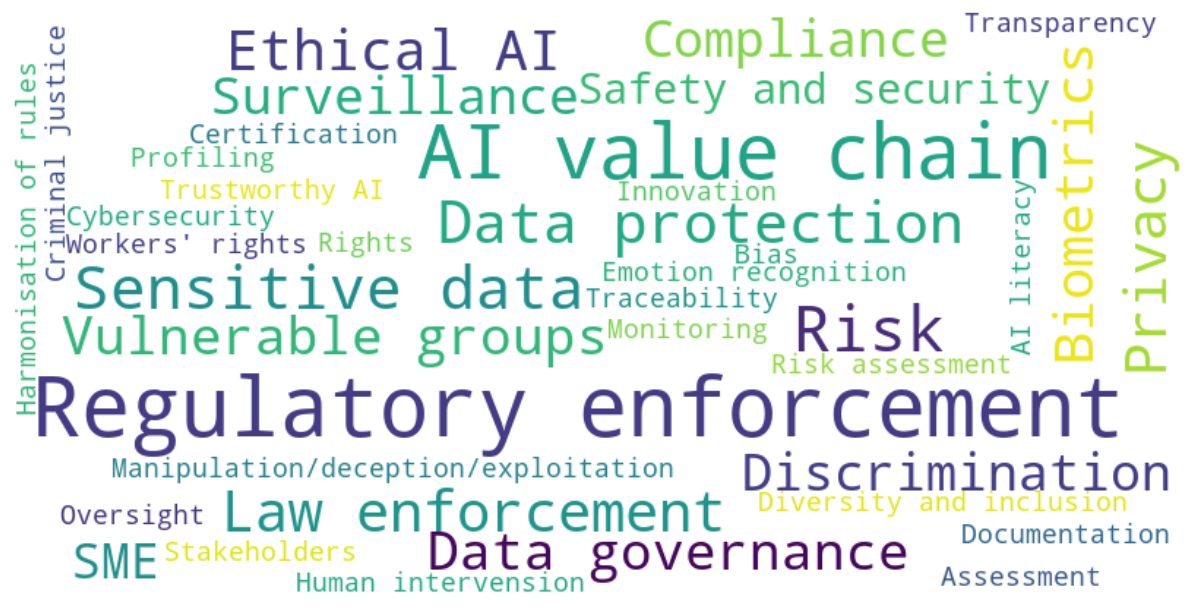}}
    \caption{Word Cloud of the identified AI governance themes from EU AI policies}
    \label{fig:<all_theme_wordcloud>}
\end{figure}

Our analysis of outputs from BERTopic was conducted manually to assign themes to each topic cluster. Despite text pre-processing and hyperparameter optimisation of  BERTopic, approximately 14\% of the clusters exhibited insufficient coherence to allow meaningful thematic interpretation. Additionally, while some topic clusters were coherent, they correspond to multiple themes. We adopted a minimalistic approach when assigning themes, i.e. attributing the smallest set of themes to each cluster. In cases where comprehensive representation required multiple themes, we established a constraint of assigning a maximum of 3 themes to each cluster. The analysis resulted in \textbf{100 distinct themes} in total. \autoref{tab:<bert_topics_themes>} shows the number of topic clusters and  themes, along with the key themes identified in each EU AI policy document. 
\autoref{fig:<all_theme_wordcloud>} provides an aggregated thematic overview using a word cloud visualisation, highlighting the most prevalent themes across the corpus.


\begin{figure}[!h]
\centering
\begin{subfigure}{0.5\textwidth}
\fbox{\includegraphics[width=\linewidth, height=6.5cm]{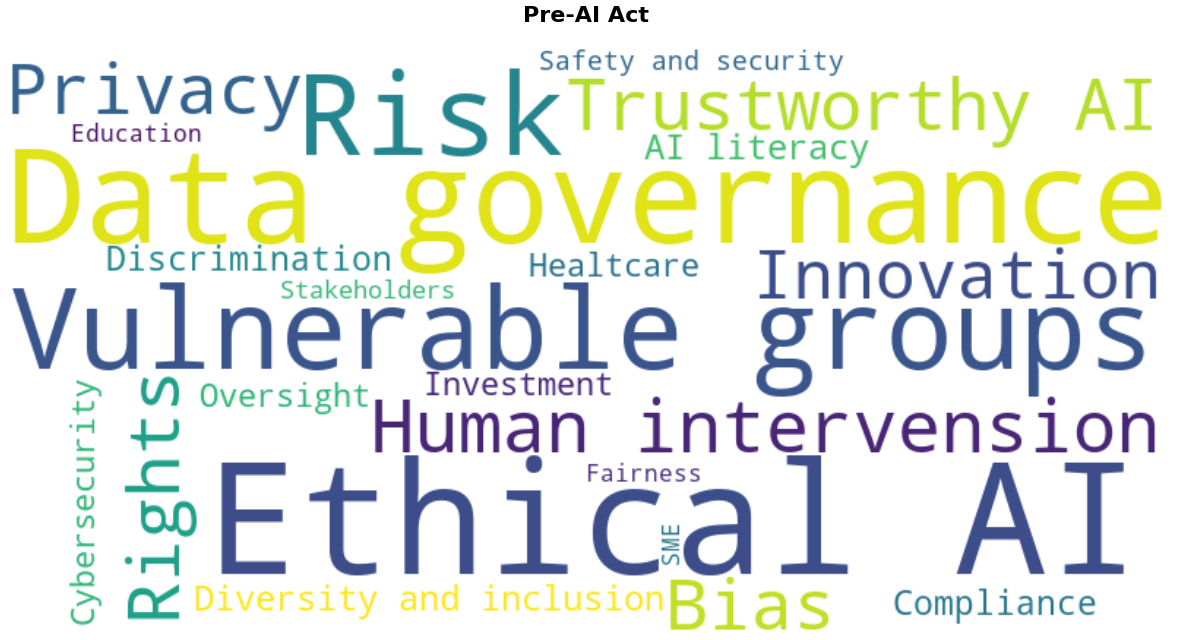}} 
\caption{Themes pre-AI Act}
\label{fig:<sub-themes-preaiact>}
\end{subfigure}%
\begin{subfigure}{0.5\textwidth}
\fbox{\includegraphics[width=\linewidth, height=6.5cm]{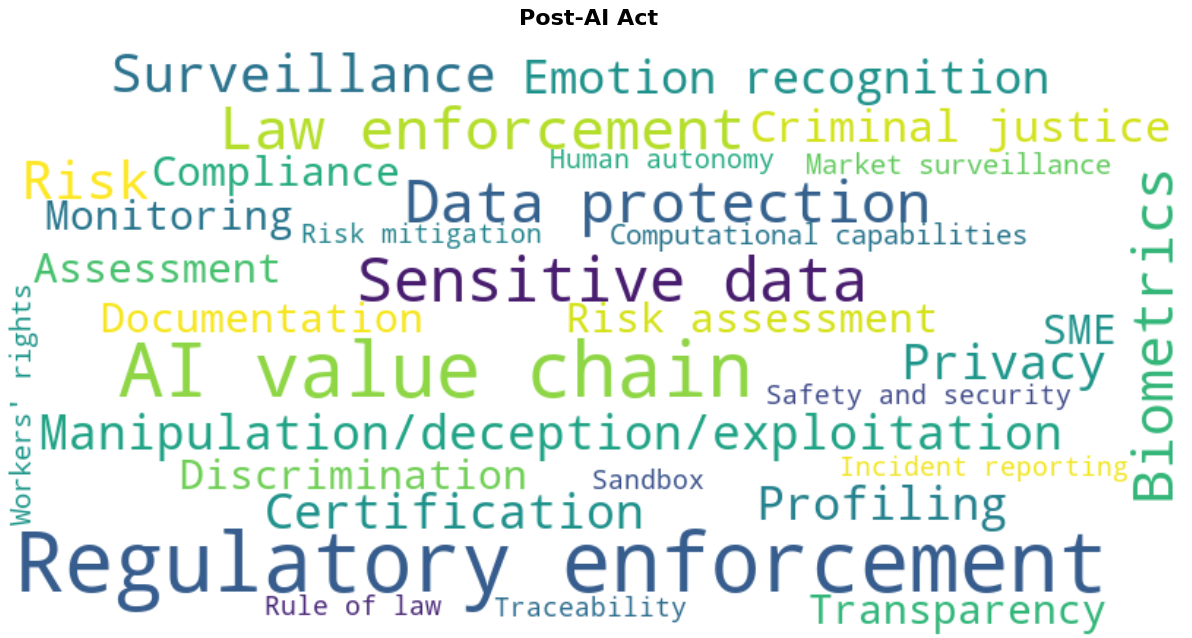}}
\caption{Themes post-AI Act}
\label{fig:<sub-themes-postaiact>}
\end{subfigure}

\caption{Word clouds demonstrating evolution of AI governance themes in the EU}
\label{fig:image2}
\end{figure}

\subsection{Evolution of Themes in EU AI Policies}

\begin{figure}[ht]
\centering

\begin{subfigure}{\textwidth}
\centering
\fbox{\includegraphics[width=\linewidth, height=4.5cm]{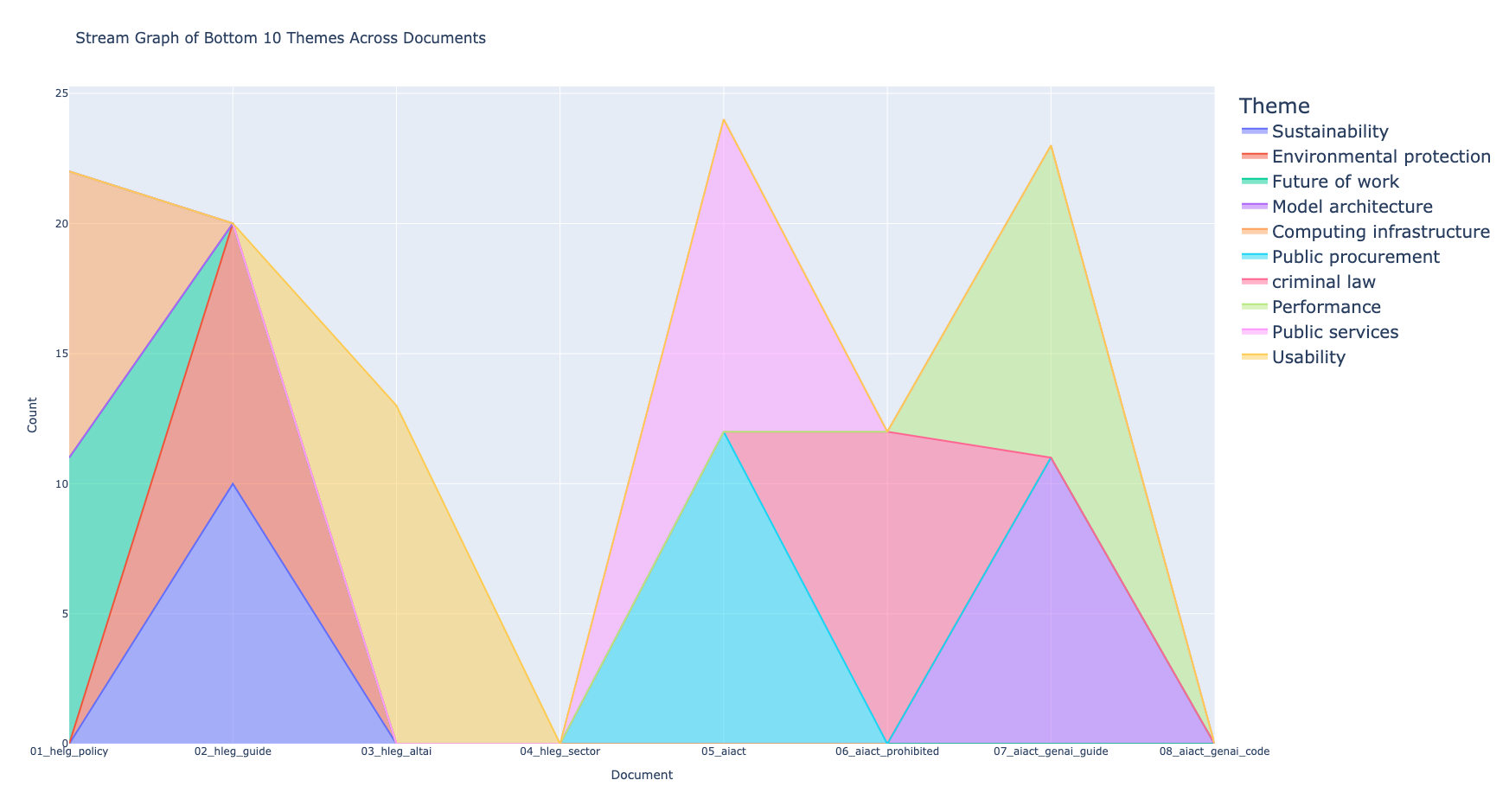}} 
\caption{Bottom 10 themes}
\label{fig:<sub-10bottom-streamgrapg>}
\end{subfigure}

\begin{subfigure}{\textwidth}
\centering
\fbox{\includegraphics[width=\linewidth, height=4.5cm]{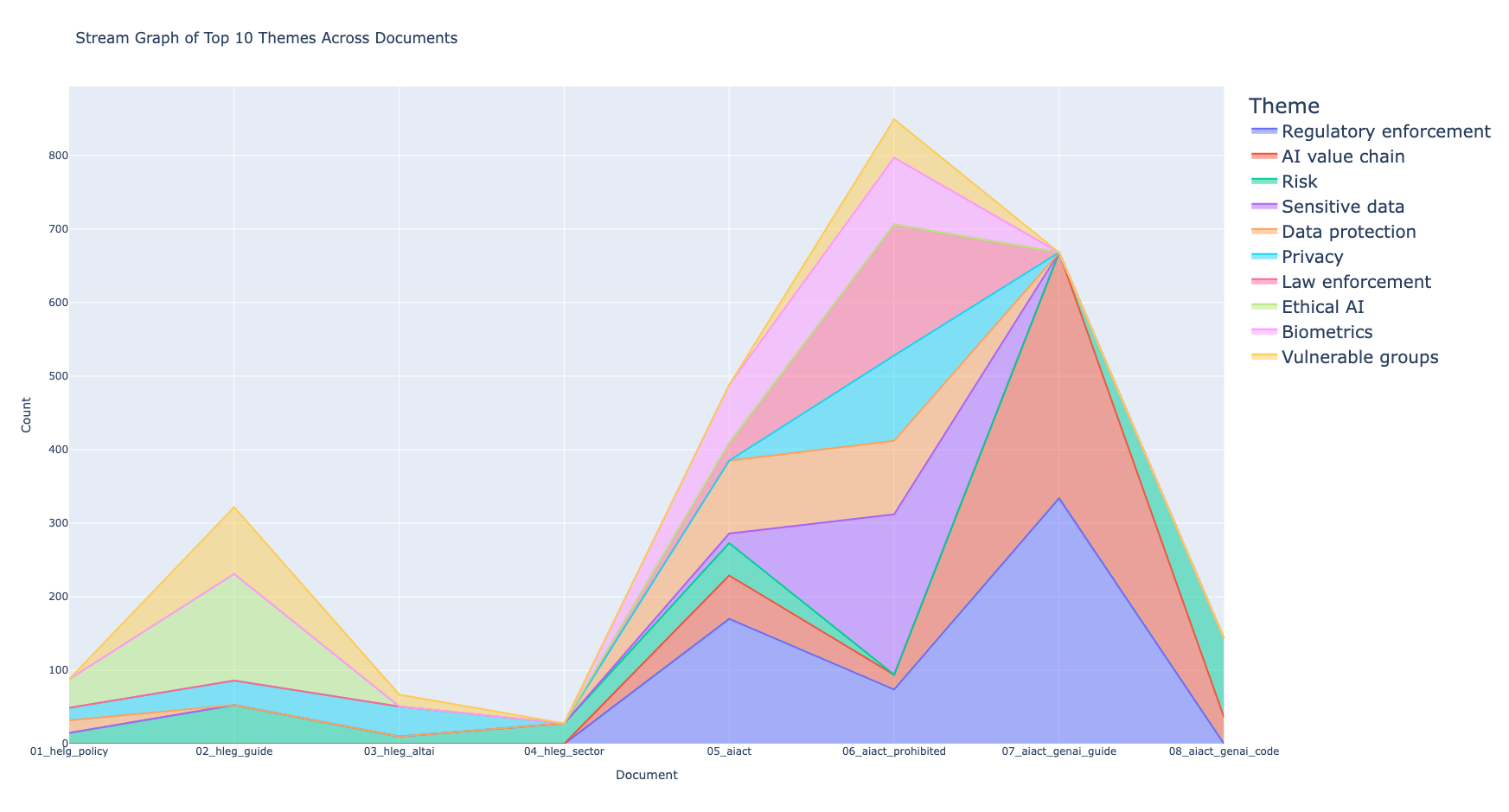}}
\caption{Top 10 themes}
\label{fig:<sub-10top_streamgraph>}
\end{subfigure}

\caption{Stream graphs demonstrating evolution of most and least prevalent topics  over time}
\label{fig:<streamgraph>}
\end{figure}

Our analysis shows a shift in the AI policy discourse in the EU following the adoption of the EU AI Act. As shown in \autoref{fig:<sub-themes-preaiact>}, the pre-AI Act thematic landscape predominately focused on \textit{ethical AI}, \textit{data governance}, and \textit{risks} particularly those that impact \textit{vulnerable groups}. This emphasis aligns with both the non-binding nature of the EU HLEG documents and their mission to promote ethical AI, raise awareness of AI risks, and provide a strategic roadmap, particularly in 2019–2020, when the concept of trustworthy AI was emerging as a popular yet still unfamiliar topic. 
Themes from the post-AI Act era (\autoref{fig:<sub-themes-postaiact>}) illustrate a clear shift from   \textit{aspirational ethical AI} toward \textit{operationalised legal AI}, with particular emphasis on  \textit{regulatory enforcement}.  
In addition to concrete compliance tasks such as \textit{documentation} and \textit{risk assessment}, the AI Act and its associated guidelines distinguish between entities that are involved in development, deployment and use of AI systems, i.e. \textit{AI value chain}, indicating the need for involvement of multiple entities to achieve trustworthy AI. Post-AI Act themes also revealed the focus on specific AI use cases that impose significant risk. The transition in the AI policy discourse is clearly illustrated in the stream graph of top 10 themes in \autoref{fig:<sub-10top_streamgraph>}.

 The emphasis on \textit{risk} and \textit{data protection} remains consistent from the pre- to post-legislation period, with the latter reflecting the influence of the EU GDPR. However, certain themes appear to be overlooked in this shift. To have a better understanding of such themes, we analysed the evolution of the bottom 10 (least mentioned) themes in \autoref{fig:<sub-10bottom-streamgrapg>}. The stream graph shows a waning of attention to \textit{environmental impacts} of AI as well as its broader effects on the \textit{future of work} in the post-AI Act era.

\begin{table}[!h]
    \centering
    \footnotesize
    \caption{Overview of the topics and themes associated with each document}
    \begin{tabularx}{\textwidth}{|p{0.2cm}|X|p{1.1cm}|p{0.8cm}|p{4.8cm}|}
\hline    
    \textbf{ID}& \textbf{Document} &  \textbf{No. of BERTopic clusters}& \textbf{No. of themes} & \textbf{Key themes} \\

\hline

 01 & HLEG policy and investment recommendations \cite{eu_hleg-policy} &  26
 &   23 
&
AI literacy, SME, Education, Data governance, Ethical AI, Investment, Innovation \\

\hline 
  02 & HLEG ethics guidelines \cite{eu-hleg-trustworthyai} &  28 
   & 27
   &
   Ethical AI, Rights, Vulnerable groups, Compliance, Risk, Fairness, Diversity and inclusion, Bias, Safety and security, Oversight, Privacy \\

\hline
   03 & HLEG ALTAI \cite{eu-hleg-altai} & 14 
    & 16 
    &
    Human intervention, Bias, Discrimination, Trustworthy AI, Privacy, Data governance, Cybersecurity, Safety and security, AI standardisation  
  \\
\hline   

   04 & HLEG Sectoral considerations on policy and investment recommendations \cite{eu-hleg-sectoral} &  
    8 
    & 12 
    & 
    Health Care, Innovation, Risk, Copyright, Sector-specific AI, Trustworthy AI, Investment\\
\hline

    05 & EU AI Act \cite{eu-ai-act} & 70 
      & 47 
      &
      Regulatory enforcement, Certification, Data protection, Monitoring, SME, Biometrics, Market surveillance, Traceability, Sandbox, AI value chain , General-purpose AI model, Harmonisation of rules, Discrimination, Documentation, Compliance, Risk, Cybersecurity, Oversight, Data governance, AI standardisation, Sector-specific AI
      \\
\hline 
     06 & AI Act's prohibited AI guideline \cite{eu_prohibited} &  
     62 
     & 33 
     &
     Sensitive data, Law enforcement, Surveillance, Privacy, Manipulation/deception/exploitation, Profiling, Criminal justice, Data protection, Emotion recognition, Biometrics, Transparency, Regulatory enforcement, Rule of law, Human autonomy, Vulnerable groups, National security, Psychological harm, Workers' rights, Social scoring  \\
\hline      
   07 & AI Act's general-purpose AI  guideline \cite{eu_genai_models}&  7 
    & 7 
    &  Regulatory enforcement, AI value chain, Computational capabilities, Model training \\
\hline
    08 & AI Act's general-purpose AI  code of practice \cite{eu_gpai_code}& 12
    &12 
    &  Risk, Risk assessment, Safety and security, Risk mitigation, AI value chain , Documentation, Compliance, Copyright  \\
\hline        
    \end{tabularx}
    
    \label{tab:<bert_topics_themes>}
\end{table}

\section{Conclusion and Future Work} \label{sec:<conclusion>}
This study has demonstrated the efficacy of employing BERTopic, validated and refined through expert interdisciplinary thematic analysis, to identify the complex and evolving landscape of AI governance within the EU. The identified themes cover a wide-range, including fundamental rights, risk management, technical capabilities, regulatory enforcement, AI literacy, support for innovation, and market investment. Our work advances the state of the art in three key ways: (1) it employs a combination of topic modelling and thematic analysis enabling scalable yet human-centric approach to analyse the evolving body of AI policies, (2) it provides a holistic and temporal view of EU AI policies, and (3) it concentrates specifically on AI governance and success criteria for the responsible development and deployment of AI.



In future work, we plan to expand our analysis and investigate AI governance themes that emerged in the guidelines set forth by supranational bodies, such as UNESCO, The Organisation for Economic Co-operation and Development (OECD), and the Council of Europe. Given the importance of technical standards in the implementation of trustworthy AI guidelines as well as the EU AI Act, we also aim to identify the themes in AI standards developed by international and European standardisation bodies, namely ISO/IEC JTC 1/SC 42 and CEN/CENELEC JTC 21. Comparison of the themes identified from these two types of AI documents will assist in clarifying the interplay between technical standards and trustworthy AI guidelines and regulations. Moreover, from a methodological perspective we aim to incorporate more refined and interpretable topic models \cite{pham2024topicgpt} leading to increased scalability across a larger set of documents. Our methodologial approach aims to optimise  human-AI collaboration in policy document analysis.


\section*{Acknowledgements}
This work has received funding from the European Commission's Horizon Europe Research and Innovation Programme under grant agreement No. 101177579 (FORSEE) and from Research Ireland  at ADAPT, the Research Ireland Centre for AI-Driven Digital Content Technology at TCD and UCD \#13/RC/2106\_P2. For the purpose of Open Access, the authors have applied a CC BY public copyright licence to any Author Accepted Manuscript version arising from this submission. We would like to thank Prof. John D. Kelleher and Dr. Vasudevan Nedumpozhimana for providing valuable advice on topic modelling. 

\bibliographystyle{vancouver}
\bibliography{references}
\end{document}